\documentclass[letterpaper]{article}

\usepackage[T1]{fontenc}

\usepackage{geometry}
\usepackage{setspace}

\usepackage[style = chem-acs, articletitle=true]{biblatex}
\usepackage{graphicx}
\usepackage{float}
\newfloat{scheme}{htbp}{los}
\floatname{scheme}{Scheme}
\floatname{chart}{Chart}
\newfloat{graph}{htbp}{loh}
\usepackage{makecell}
\usepackage{chemformula} 
\usepackage[version = 4]{mhchem} 
\usepackage{hyperref}
\usepackage{bm}

\usepackage{authblk}
\author[1]{Dahvyd Wing}
\author[1]{Alexandre Tkatchenko*}
\affil[1]{Department of Physics and Materials Science, University of Luxembourg, L-1511 Luxembourg City, Luxembourg}

\title{A Transferable Model of Molecular Exchange–Repulsion Interaction from Anisotropic Valence Density Overlap}

\date{*Email: alexandre.tkatchenko@uni.lu}

\begin{document}

\maketitle

\begin{abstract}
Pauli exchange-repulsion is the dominant short-range intermolecular interaction and it is an essential component of molecular force fields. Current approaches to modeling Pauli repulsion in molecular force fields often rely on over 20 atom types to achieve chemical accuracy, illustrating the challenge in finding models which have broadly transferable parameters, which hampers the development of force fields with quantum-chemical accuracy that are transferable across many chemical systems. We present the anisotropic valence density overlap (AVDO) model for exchange-repulsion. The model produces sub-kcal/mol accuracy for dimers of organic molecules and contains two universal parameters, which we demonstrate are transferable for molecules composed of H, C, N, O, F, P, S, Cl, and Br. The model is tested on 1,872 unique molecular pairs selected from a set of 135 molecules, and samples dissociation curves and configurations from condensed-phase molecular dynamics trajectories. Given recent progress in machine learning of the electronic density, this model offers a promising path toward high-accuracy, next-generation machine-learned force fields.
\end{abstract}






\section{Introduction}

Classical force fields of organic molecules have been indispensable in molecular biology, drug design, crystal structure prediction, and many other fields \cite{dauber2019biomolecular,hagler2019force}. A model for Pauli exchange-repulsion, the short-range repulsion between atoms and molecules due to electrons obeying the Pauli exclusion principle, is a critical component of these force fields. Many macroscopic properties of condensed molecular systems, such as the density, are determined by the balance between exchange-repulsion and attractive intermolecular interactions \cite{stone_2013_book}. It is also a key factor in determining the stacking conformation of $\pi-\pi$ interactions in proteins \cite{carter2023energetic} and the preference of linear conformations of halogen sigma bonds \cite{stone2013halogen}, which are prevalent in drug design because they increase drug specificity \cite{xu2014halogen}. 

Recently, intermolecular potentials have been parameterized using the components of symmetry-adapted perturbation theory (SAPT) \cite{mcdaniel2013transfer, van_vleet_2016_exch, vandenbrande2017monomer, liu2019amoeba+, schriber2021cliff, rackers2021hippo}, a method that calculates dimer interaction energies as the sum of exchange-repulsion, electrostatic, induction, and dispersion interactions \cite{jeziorski1994sapt_review, stone_2013_book}. In benchmark dimer datasets, SAPT shows that exchange-repulsion is often the single largest energy component and has the largest variance \cite{szalewicz2022physical, parker2014levels,liu2019amoeba+, sparrow2021nenci}. Thus, it is not surprising that recent intermolecular potentials have found that the largest source of error in modern force fields is the exchange-repulsion term \cite{van_vleet_2016_exch, vandenbrande2017monomer, liu2019amoeba+, schriber2021cliff} and there has been a renewed interest in developing accurate exchange-repulsion models beyond Lennard-Jones \cite{lennard_jone_1931} and Born-Mayer potentials \cite{born_mayer_1932, buckingham_1938} \cite{rackers_2019_S2R, chung_ponder_2024_S2R_p_orb}. 

In an effort to reduce the error of recently developed exchange-repulsion models, tens of atom-type parameters have been used \cite{rackers_2019_S2R, schriber2021cliff} and this limits transferability to molecules composed of these atom-types. Additionally, transferability has generally been assessed using near-equilibrium configurations \cite{schriber2021cliff}, leaving open whether these models retain their accuracy for non-equilibrium configurations relevant to molecular dynamics. This work examines transferability across both chemical space and configuration space.

For simple Lennard-Jones or Born-Mayer  models for repulsion, it is inevitable that one must employ many atom-type parameters because these models are isotropic, whereas an atom-centered model of exchange repulsion should be anisotropic \cite{stone_price1988aniso, wheatley_1990_exch, day2003nonempirical,totton2010, stone_2013_book,elking2010gmm, van_vleet2018mastiff, rackers_2019_S2R, chung2024_hippo_aniso}, and they have the wrong distance dependence \cite{van_vleet_2016_exch}. Thus, they must be fit per molecule and often rely on error cancellation with other terms in the force field to achieve accuracy~\cite{van_vleet_2016_exch}.

The density overlap model~\cite{kita1976, kim_1981_ovlp}, an empirical model, calculates the exchange-repulsion energy between two molecules as
\begin{equation}
\label{eq:ovlp_model}
E_{\textrm{exch}} =K \bigg(\int \rho_A \rho_B \textrm{d}^3r\bigg)^\alpha,
\end{equation}
where $K$ and $\alpha$ are fitted parameters and $\rho_A$ and $\rho_B$ are the electron densities of molecules $A$ and $B$. Density overlap models have the potential to significantly reduce the number of free parameters that must be fit against SAPT exchange-repulsion energies, because the density largely determines the distance dependence and angular dependence of the model, and thus these models are under active development \cite{van_vleet_2016_exch, vandenbrande2017monomer, van_vleet2018mastiff, 2021_cisneros_rep_dens, schriber2021cliff, 2023_van_vleet_benzene, 2024_cisneros_lichem}. However, there is no single universal $K$ parameter that yields sub-kcal/mol accuracy for all molecules \cite{gavezzotti2003, soderhjelm_2006_review} and instead $K$ must be fit per dimer pair \cite{piquemal_darden_2006, elking2010gmm, bygrave_manby2012,vandenbrande2017monomer}, with values varying by 20–30\% \cite{kim_1981_ovlp,  piquemal_darden_2006, elking2010gmm, ihm_1990_charge, nobeli1998_different_K}. To increase transferability, the density can be separated into atom-centered partial densities \cite{lillestolen_2009_isa, verstraelen_2016_mbis} so that different atom-type parameters (typically 20-30 atom types) can be used with simple combination rules \cite{mitchell_price2000atomtypes, van_vleet_2016_exch, schriber2021cliff}. However, the fact that the parameter for a single element can vary by as much as a factor of 2 suggests that transferability is limited to similar molecules, or even that every molecule must be fit individually \cite{van_vleet_2016_exch, chung_ponder_2024_S2R_p_orb}. 

Advances in machine learning (ML) have now opened up new considerations for the development of density overlap methods. In the past, density overlap models employed simplified, often isotropic, model electron densities \cite{elking2010gmm, vandenbrande2017monomer, van_vleet_2016_exch, schriber2021cliff}. This was due to the strong constraint that the models needed to be extremely computationally efficient for application in long-time molecular dynamics (MD) and high throughput docking studies and because there existed ML methods that could predict these simplified densities \cite{bereau2018_ipml, schriber2021cliff}. With the advent of machine-learned force fields \cite{unke2021mlff}, there is now a need for large datasets of high quality synthetic data to train machine-learned force fields \cite{kulichenko2024active_learning}. This is a less computationally demanding task, such that density overlap models which use the full electron density can be considered. Additionally, advanced ML models have now been developed that can predict the full electron density \cite{brockherde2017_ml,bogojeski2018_ml_dens, grisafi2018_transferable_ml_dens, fabrizio2019_dimer_ml_dens}.

Thus, it is timely to reassess the design space of density overlap models. Specifically, being able to use the full, anisotropic electron density opens another degree of freedom that has not been deeply explored, namely that one can remove density from lower energy orbitals in a density overlap model. It has been observed that removing core electrons does not significantly change the exchange-repulsion potential energy surface (PES) for water \cite{tafipolsky2016} and recent models implicitly neglect core electrons \cite{van_vleet_2016_exch, vandenbrande2017monomer, rackers_2019_S2R}. Since density functional theory (DFT) molecular orbitals decay as $\sim e^{-\sqrt{-2\epsilon} r}$, where $\epsilon$ is the energy of the molecular orbital \cite{misquitta2016pyridine, 2016asymptotic_dft} it is clear that core orbitals will have a smaller contribution to the density overlap. 

In this article, we show that by using density overlap models on densities from valence molecular orbitals, one can improve the accuracy of exchange-repulsion predictions by nearly a factor of two. The resulting model uses only two universal parameters and is transferable across small organic molecules containing H, C, N, O, F, P, S, Cl, and Br, achieving sub-kcal/mol accuracy. We further evaluate the model on non-equilibrium dimer configurations relevant to molecular dynamics and show that this accuracy extends beyond near-equilibrium geometries.

\section{Methods}
We investigate the accuracy and transferability of density overlap models that use partial densities that systematically exclude contributions from low-energy molecular orbitals. Partial densities are computed by:
\[\rho_{\textrm{partial}} = \sum_{i=M+1}^{N} \big|\psi_i\big|^2,\]
where $N$ is the total number of electrons in the monomer, $\psi$ are molecular orbitals, and $M$ is the number of orbitals that are not included in the density. $M$ is calculated by:
\begin{equation}
\label{eq:M_low_orbitals}
M = \sum_{X \in \{\textrm{C, N, O, ... }\}} M_{X} n_X,
\end{equation}
where $n_X$ is the number of atoms of element $X$ in the molecule; and $M_{X}$ are the number of molecular orbitals to neglect per atom of element $X$. This heuristic method removes low-energy molecular orbitals associated with core and semi-core atomic orbitals. The $M_X$ parameters define the partial density and we denote each partial density by the following notation: C$^{M_C}$N$^{M_N}$O$^{M_O}$F$^{M_F}$..., with $M_H$ always set to 0. In this work, monomer partial densities are computed using DFT with the PBE functional \cite{pbe}.

For each of these partial densities, we compute density overlaps analytically using the monomer densities. We then fit the parameters of the density overlap model (eq. \ref{eq:ovlp_model}) by minimizing the mean absolute error with respect to reference exchange-repulsion energies, which we compute using SAPT(DFT) \cite{williams2001sapt_dft, misquitta2002sapt_dft} with the PBE functional. Choosing the mean absolute error as the cost function reduces the influence of highly repulsive configurations on the fit. The fitting process uses the S101x7 dataset \cite{wang2015_penetration_s101}. This dataset contains 101 dimers each sampled at 7 intermolecular distances along a dissociation curve, though we exclude configurations with a separation of 0.7 $R_{eq}$, where $R_{eq}$ is the equilibrium dimer distance, because these configurations have an average interaction energy of 30 kcal/mol above the minima and are therefore not accessible at room temperature. The closest configurations in the dataset, the configurations at $0.8R_{eq}$, have an average interaction energy of 8 kcal/mol above the minima.

The fitting procedure is done in three steps. First, we fit linear models ($\alpha =1$ and only $K$ is fit) and power-law models (both $K$ and $\alpha$ are fit simultaneously) for each partial density on the S66 subset of S101x7 \cite{s66}. The S66 subset contains 66 combinations of 14 neutral, closed-shell molecules composed of hydrogen, carbon, nitrogen, and oxygen, including small molecules such as water, ethyne, pentane, and benzene and biologically relevant molecules like N-methylacetamide and uracil. The dimer configurations in the dataset sample the most common noncovalent interactions in biomolecules. This step determines the optimal $\alpha$ and $K$ for a given set of $M_C$, $M_N$, and $M_O$. In the second step, we freeze $K$ and $\alpha$ and search for partial densities that are the most accurate on a set of 23 neutral dimers of S101x7 which contain the elements F, P, S, Cl, and Br, which we call the S101-FPSClBr dataset. This determines the optimal $M_F$, $M_P$, $M_S$, $M_{Cl}$, and $M_{Br}$, for a given set of $K$, $\alpha$, $M_C$, $M_N$, and $M_O$ parameters. Third, we select the model and partial density that gives the best overall accuracy on the combined S66 and S101-FPSClBr datasets.

We then test the transferability of the best model on neutral molecules from the DES15K dataset \cite{DES15K}. This dataset contains two parts. The first part consists of monomers in their relaxed configuration along a dissociation path starting from an optimized dimer configuration. This yields 4,063 configurations consisting of 1,016 unique dimers of 78 different molecules composed of H, C, N, and O and 3,423 configurations of 856 dimers which include 53 molecules containing F, P, S, Cl, and Br. Each dimer is sampled by at most 4 points along a dissociation curve. The second part of the dataset contains out-of-equilibrium geometries taken from nearest neighbor dimers from condensed phase MD simulations at 298 K. This MD dataset is further divided into a dataset containing molecules that are composed of H, C, N, and O, consisting of 74 molecules solvated in water and 56 molecules in a neat liquid (1,580 configurations in total); and a dataset of molecules that contain F, P, S, Cl, and Br, consisting of 37 molecules solvated in water and 14 molecules in a neat liquid (473 configurations in total). For a list of the classes of compounds contained in these two datasets, see Table \ref{tbl:AVDO_by_group}. The intermolecular distance between the closest two atoms of a dimer goes from 1.1 to 4.7 angstroms for the optimized dissociation curve dataset and from 1.5 to 3.3 angstroms for the MD dataset. 

\subsection{Computational details}
Densities and overlaps are computed using \texttt{PySCF} \cite{2018pyscf, 2020pyscf, 2015libcint} with an aug-cc-pVTZ Gaussian dimer basis set, i.e. ghost atoms are placed at the positions of the other monomer. Density overlaps are converged to less than approximately 1\% for uracil and 0.01 \% for water using this basis. This results in model errors around 0.3 kcal/mol at equilibrium for the uracil dimer.

SAPT(DFT) calculations are performed in \texttt{Q-Chem} \cite{2021qchem} using a smaller, aug-cc-pVDZ, Gaussian dimer basis set due to computational cost. SAPT(DFT) exchange-repulsion energies are converged to better than 0.2 kcal/mol for the uracil dimer at 0.8$R_{eq}$, with errors that decrease with distance such that they are negligible at the equilibrium distance and beyond. For the acene series, SAPT(DFT) exchange-repulsion energies were calculated using \texttt{PySCF}. 

The Nelder-Mead algorithm as implemented in \texttt{SciPy} \cite{2020SciPy} is used to minimize the mean absolute error in the exchange-repulsion energy for a simultaneous fit of $K$ and $\alpha$. Due to the fact that $\alpha$ is close to one, it is possible to choose good initial values for $K$ and $\alpha$ such that the fitting procedure is robust. 

\section{Results}
\begin{figure}[htbp]
    \centering
    \includegraphics[width=.36\textwidth]{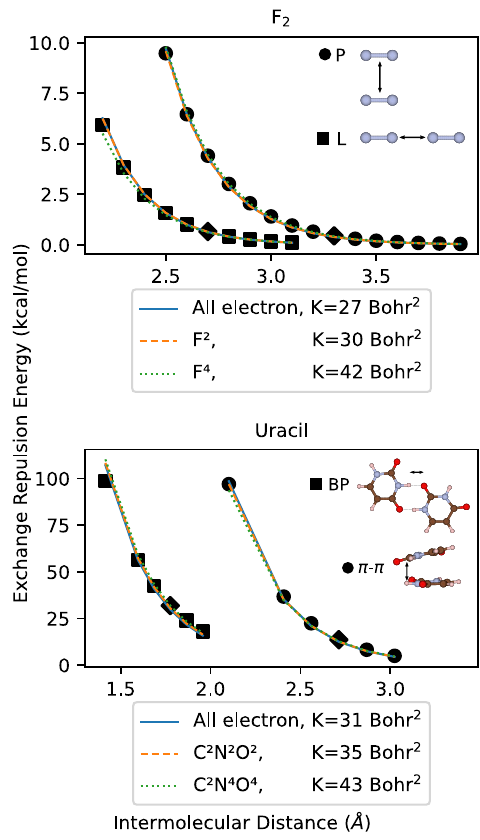}
    \hfill
    \caption{Linear density overlap models (lines) fitted individually for the F$_2$ dimer and the uracil dimer. Squares and circles are the SAPT(DFT) exchange repulsion energies for the parallel or $\pi-\pi$ stacking dimer configurations (circles) and the linear or base-pair dimer configurations (squares). Diamonds mark the minima of the total interaction energy along that trajectory. The models used the all-electron density, the density without orbitals from core electrons (F$^2$ and C$^2$N$^2$O$^2$), or the density without orbitals corresponding to core electrons and semicore electrons (F$^4$ and C$^2$N$^4$O$^4$).  }
    \label{fig:F2_uracil_1}
\end{figure}

First, we illustrate the general behavior of partial density overlap models by examining F$_2$ and uracil homodimers. We use linear models for this case and we fit $K$ separately for each dimer and partial density. For each fit, we use two dimer configurations sampled at several intermolecular distances. The results are presented in Fig. \ref{fig:F2_uracil_1}.

We observe that the exchange-repulsion PES changes only slightly for density overlap models using different partial densities. Removing low-energy orbitals for the F$^4$ and C$^2$N$^2$O$^4$ models reduced density overlap by 30\%, which was compensated for by the commensurate increase in the $K$ parameter. Importantly, while the resulting PES remains largely unaltered, we observe that the difference between the fitted $K$ parameters of the two dimers decreases by using these partial valence densities. This illustrates the possibility that using a partial density can increase the transferability of the density overlap model.

\begin{figure}[htbp]
    \centering
        \includegraphics[width=0.36\textwidth]{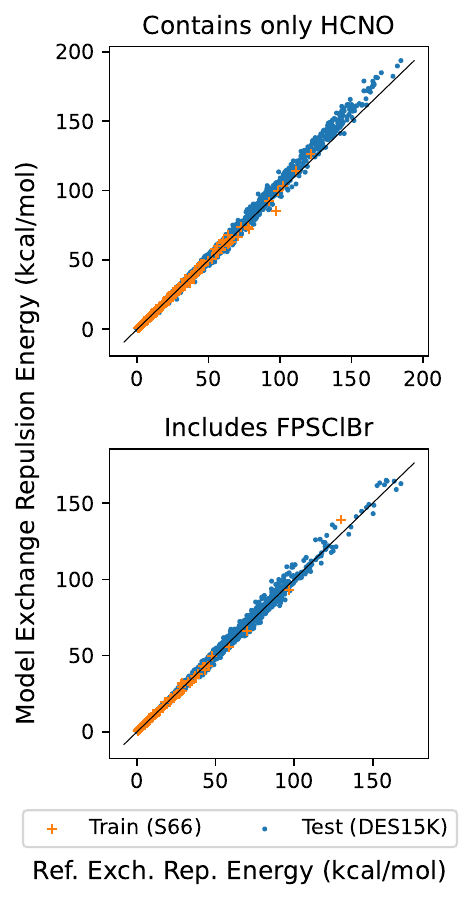}
    \hfill
    \caption{Exchange repulsion energies (kcal/mol) for the AVDO power-law model trained on the S66 dataset and the S101x7 FPSClBr dataset, and tested on the DES15K dataset. The black line represents perfect agreement between the reference and the model.}
    \label{fig:compare_all}
\end{figure}

The results of fitting different partial density overlap models parameterized on the S66 subset of S101x7 are presented in Table \ref{tbl:compare_models} and Fig. \ref{fig:compare_all}. We see that the models which remove the core electrons (C$^2$N$^2$O$^2$), the core electrons and the 2s electrons of oxygen (C$^2$N$^2$O$^4$), and core electrons and the 2s electrons of oxygen and nitrogen (C$^2$N$^4$O$^4$) all have lower root mean square errors (RMSE) than the all-electron density model, both for linear models and for power-law models. The power-law models are more accurate than the linear models at all ranges, with the greatest improvement at shorter distances. The power-law model with the C$^2$N$^4$O$^4$ partial density is nearly twice as accurate as the power-law all-electron model and has a relative RMSE of 5\% for equilibrium geometries with respect to the standard deviation (STD) of the reference SAPT(DFT) exchange-repulsion energies.

  \begin{table}
  \caption{RMSE of Exchange Repulsion Models Using Different Partial Densities on the S66 dataset (kcal/mol)}
  \label{tbl:compare_models}
\centering
  \begin{tabular}{lccccc}
  \hline
Model    & \shortstack{ 0.8 $R_{eq}$ }
     & \shortstack{ 0.9 $R_{eq}$ }
    & \shortstack{ 1.0 $R_{eq}$ }
    & \shortstack{1.1 $R_{eq}$ }& \makecell{Parameters$^a$ \\($K$, $\alpha$)}\\ 
    \hline
    Linear Models \\
All-electron             & 6.6  & 2.4  & 0.9 &  0.5   & (32.4, 1)\\ 
C$^2$N$^2$O$^2$          & 3.8  & 1.2  & 0.7 &  0.5   & (38.3, 1)\\
C$^2$N$^2$O$^4$          & 3.7  & 1.2  & 0.6 &  0.5    & (38.6, 1)\\
C$^2$N$^4$O$^4$          & 3.6  & 1.1   & 0.6   & 0.5  & (41.0, 1)\\
C$^4$N$^4$O$^4$          & 7.9  & 4.2   & 2.3   & 1.3  & (58.6, 1)\\
\hline
Power-law Models\\
All-electron             & 4.0  & 1.5  & 0.7 &  0.4    & (18.9, 0.92)\\ 
C$^2$N$^2$O$^2$          & \bf{2.1}  & 0.9  & 0.5 &  0.3   & (26.8, 0.95) \\
C$^2$N$^2$O$^4$          & 2.4  & 0.9  & 0.4 &  0.3    & (28.8, 0.96)\\
\bf{C$^2$N$^4$O$^4$}$^b$     & 2.4  & \bf{0.8}   &\bf{0.4}   & \bf{0.3}& (30.4, 0.96) \\
C$^4$N$^4$O$^4$          & 8.0  & 4.2   & 2.3   & 1.3   & (55.4, 0.99)\\
Ref. mean $\pm$ STD$^c$ & 46.0 $\pm$ 23.9   & 20.6 $\pm$ 13.0   &  9.5  $\pm$ 8.0  &  4.6  $\pm$  4.9    \\
\hline
    \end{tabular}

      \parbox{\linewidth}{\vspace{1ex} \centering  \small a) $K$ is in Hartree and the overlap is in Bohr$^{-3}$  b) AVDO model c) The mean of the reference data $\pm$ the standard deviation of the reference data}
  \end{table}

Since the C$^2$N$^2$O$^2$, C$^2$N$^2$O$^4$, and C$^2$N$^4$O$^4$ partial densities all yield similar results, we extend each of these partial densities for molecules that contain the elements F, P, S, Cl, and Br using the S101-FPSClBr dataset. We keep the $\alpha$ and $K$ parameters fixed for each of these partial densities (see Table \ref{tbl:compare_models} for values) and test different values of $M_X$ in eq. \ref{eq:M_low_orbitals} for F, P, S, Cl, and Br. The best models using extended valence densities are shown in Table \ref{tbl:FPSClBr_compare_models} (see the SI for more details). In general, it is found that removing only a fraction of an S or P block of semicore states yields poor results. This matches chemical intuition and dramatically reduces the parameter space that needs to be searched. From Table \ref{tbl:FPSClBr_compare_models} we determine that the power-law model using a C$^2$N$^4$O$^4$F$^4$P$^{10}$S$^{10}$Cl$^{10}$Br$^{28}$ valence density yields the best accuracy overall. We call this model the anisotropic, valence density overlap (AVDO) model.

    \begin{table}
  \caption{RMSE of Exchange Repulsion Models using different partial densities on the S101-FPSClBr dataset (kcal/mol)}
  \label{tbl:FPSClBr_compare_models}
  \centering
  \begin{tabular}{lccccc}
  \hline
 Model
    & 0.8 $R_{eq}$ &  0.9 $R_{eq}$  
    & 1.0 $R_{eq}$  & 1.1 $R_{eq}$ \\
    Linear Models \\
All-electron                                             & 7.3 & 2.1 & 0.7 & \bf{0.2}  \\
C$^2$N$^2$O$^2$F$^2$P$^{12}$S$^{12}$Cl$^{4}$Br$^{22}$    & 6.5 & 2.5 & 1.0 & 0.4  \\
C$^2$N$^2$O$^4$F$^4$P$^{4\hphantom{0}}$S$^{4\hphantom{0}}$Cl$^{4}$Br$^{18}$      & 8.0 & 2.3 & 0.8 & 0.3 \\
C$^2$N$^4$O$^4$F$^4$P$^{10}$S$^{10}$Cl$^{4}$Br$^{22}$    & 4.7 & \bf{0.9} & 0.6 & 0.5 \\

\hline
Power-law models \\
All-electron                                             & 3.4 & 1.4 & 0.7 & 0.4 \\
C$^2$N$^2$O$^2$F$^2$P$^{12}$S$^{12}$Cl$^{10}$Br$^{22}$   & 4.7 & 2.1 & 1.1 & 0.6 \\
C$^2$N$^2$O$^4$F$^4$P$^{4\hphantom{0}}$S$^{10}$Cl$^{4\hphantom{0}}$Br$^{22}$     & 3.4 & 1.5 & 0.7 & 0.3 \\
\bf{C$^2$N$^4$O$^4$F$^4$P$^{10}$S$^{10}$Cl$^{10}$Br$^{28}$}$^a$   & \bf{2.6} & 1.0 & \bf{0.6} & 0.4 \\
Ref. mean $\pm$ STD$^b$ & 35.2 $\pm$ 27.9  & 15.0 $\pm$ 14.4  & 6.8 $\pm$ 8.4  & 3.3 $\pm$ 5.1\\ 
\hline
    \end{tabular}
       \parbox{\linewidth}{\vspace{1ex} \centering \small a) AVDO model b) The mean of the reference data $\pm$ the standard deviation of the reference data}

  \end{table}

The AVDO model's accuracy on the S101-FPSClBr dataset is only slightly lower than on the S66 dataset, suggesting that the model is transferable. Thus, we test its transferability on a much larger dataset, DES15K. The results are shown in Table \ref{tbl:DES15K} and Fig. \ref{fig:compare_all}. We see that the power-law model is again more accurate than the linear model. The repulsive wall configurations in DES15K are generated by decreasing the intermolecular distance of the dimer until the total energy is 3$E_{eq}$ higher in energy than $E_{eq}$, the equilibrium binding energy along the dissociation path. In general, this places the dimer closer than the $0.8R_{eq}$ configurations of S66. The short-range configurations are at the zero-energy crossing point, which places these configurations between $0.8R_{eq}$ and $0.9R_{eq}$. Finally, the medium-range configurations are $0.5E_{eq}$ above the minima, which places them well past $1.1R_{eq}$. Thus, only the equilibrium configurations of the DES15K set can be directly compared with the S66 and S101-FPSClBr datasets. We see that the accuracy of the AVDO model on the DES15K datasets, 0.5 kcal/mol RMSE, remains the same as its performance on the S101x7 dataset, showing that the model is indeed transferable for this significantly larger dataset of organic molecules. This is the core result of this study.

  \begin{table}
  \caption{RMSE of Exchange-Repulsion Models on the DES15K dataset (kcal/mol)}
  \label{tbl:DES15K}
  \centering
  \begin{tabular}{lcccccc}
  \hline
  Model
    & \makecell[c]{ Repulsive \\[-2pt] wall }
     & \makecell[c]{ Short \\[-2pt] range }
    & \makecell[c]{ Equilibrium }
    & \makecell[c]{ Medium \\ [-2pt] range }
    &  \makecell[c]{MD \\[-2pt] configurations} 
    & ($K$, $\alpha$)$^a$ \\
    \hline
\multicolumn{7}{c}{H, C, N, and O dataset}\\
\underline{Linear Model}&&&&&& \\
AVDO fit on S66  &  11.1 &   3.5 &   0.6 &   0.2 & 0.5& (41.0, 1) \\
\underline{Power-law Models}&&&&&& \\
AVDO fit on S66    &   5.3 &   1.9 &   0.5 &   0.1 & 0.3& (30.4, 0.96) \\
AE$^b$ fit on DES15K &   5.5 &   2.5 &   0.8 &   0.2 & 0.9& (13.8, 0.87)\\
AVDO fit on DES15K    &   3.0 &   1.6 &   0.5 &   0.1 & 0.3& (26.2, 0.94) \\
Ref. mean $\pm$ STD$^c$  &76 $\pm$  39 &  41 $\pm$  22 &  11 $\pm$   7 &   0.9 $\pm$   0.9 & 6.3 $\pm$   5.6 &\\
\hline
 \multicolumn{7}{c}{F, P, S, Cl, and Br dataset} \\
\underline{Linear Model}&&&&&& \\
AVDO fit on S66    &   5.6 &   1.8 &   0.6 &   0.1 & 0.5 &\\
\underline{Power-law Model}&&&&&& \\
AVDO fit on S66 &   3.0 &   1.5 &   0.5 &   0.1 & 0.4 &\\
AE$^b$ fit on DES15K    &   6.9 &   2.9 &   0.5 &   0.2 & 0.7 &\\
AVDO fit on DES15K &   3.5 &   1.9 &   0.5 &   0.1 & 0.4 &\\
Ref. mean $\pm$ STD$^c$ & 67 $\pm$  24 &  36 $\pm$  15 &   9 $\pm$  5 &   0.7 $\pm$   0.8 &4.8 $\pm$   4.1 &\\
\hline
    \end{tabular}
 a) The parameters of the density overlap model: $K$ is in Hartree and the overlap is in Bohr$^{-3}$, and the calculations for the FPSClBr dataset use the same $K$ and $\alpha$ as in the HCNO dataset. b) All-electron c) The mean of the reference data $\pm$ the standard deviation of the reference data
  \end{table}

\begin{figure}[htbp]
    \centering
        \includegraphics[width=0.6\textwidth]{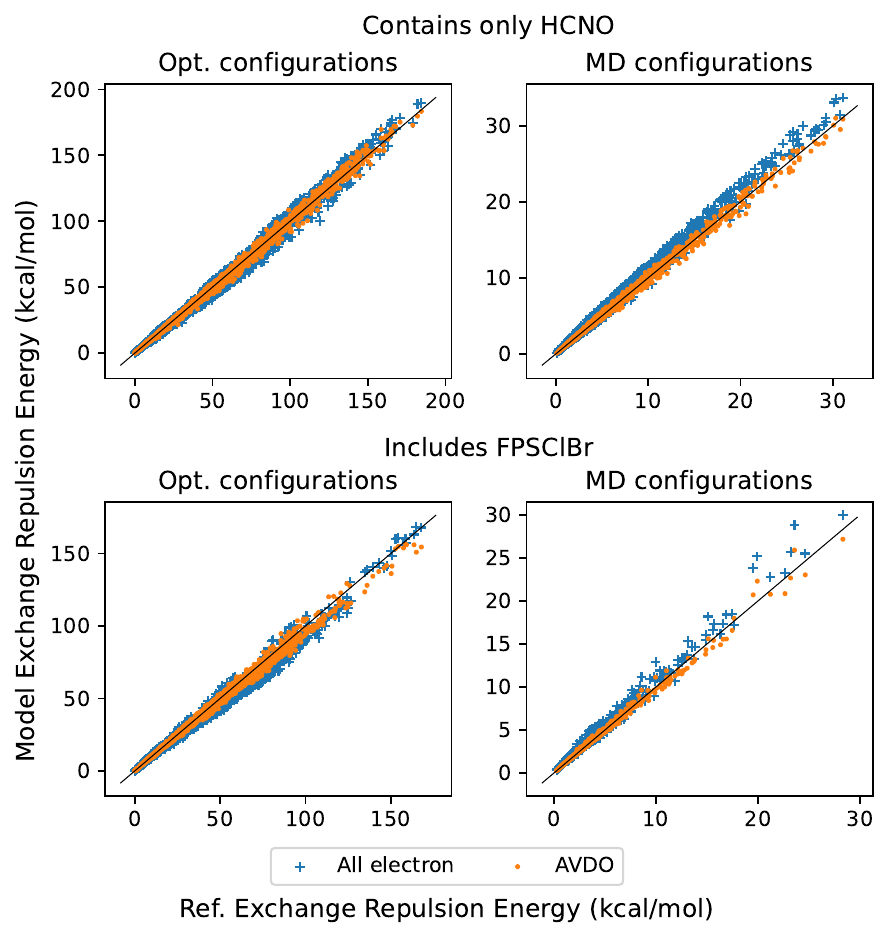}
    \hfill
    \caption{Exchange repulsion energies (kcal/mol) for the all-electron power-law model and the AVDO power-law model. Each model is only fit to the DES15K-HCNO optimized geometry dataset (the plot on top left corner). The transferability of the models to molecules that contain elements that they were not fitted on (F, P, S, Cl, and Br) is tested in the bottom two plots. Transferability to non-equilibrium configurations taken from MD trajectories at 300K is tested by the two plots on the right. The black line represents perfect agreement with the reference dataset.}
    \label{fig:DES15K}
\end{figure}

We refit both the AVDO and all-electron (AE) models on the DES15K-HCNO optimized geometry dataset to test the maximum accuracy achievable by these models on the DES15K dataset, see Table \ref{tbl:DES15K} and Fig. \ref{fig:DES15K}. By comparing to the model parameterized on S66, we see that S66 did not sample close enough configurations, such that refitting on DES15K improved the accuracy for the repulsive wall and short-range configurations. Importantly, this did not degrade the performance of the model for the DES15K-FPSClBr dataset. By comparing the all-electron model fit on DES15K and the AVDO model we again see nearly a factor of two improvement in accuracy and, importantly, an even larger improvement for non-equilibrium MD configurations. 

Further analyzing the results, we see in Table \ref{tbl:DES15K} that the error for all the density overlap models increases as the intermolecular distance decreases. Although density overlap models have the same exponential dependence as the exchange repulsion energy, they have the wrong polynomial prefactor \cite{misquitta2016pyridine, rackers_2019_S2R} and thus will not be accurate on the entire domain of intermolecular distances (see the SI for tests using Boltzmann weighting in the fitting procedure to compensate for this).

In Table \ref{tbl:AVDO_by_group}, we report the RMSE of the AVDO power-law model fit on DES15K where calculations are grouped according to whether at least one monomer in the dimer belongs to a given class of organic compounds. We find that organophosphates have the worst errors, followed by nitriles. The AVDO model for these groups has a large systematic underprediction of the repulsion energy (see the SI) showing that the optimal $K$ value for these compounds does not match the optimal $K$ value of other compounds. The top-neglected molecular orbital for the nitrile group is the sigma bond between carbon and nitrogen, which is quite exposed to neighboring molecules. Additionally, the phosphate with the most side groups has a much smaller error, probably owing to the phosphorus atom being less accessible. Analyzing the RMSE by classes of compounds also reveals that the reason the AVDO model is significantly more accurate than the all-electron model in predicting MD configurations in Table \ref{tbl:DES15K} is that the AVDO model is more accurate than the all-electron model for dimers containing water, and more than half of the MD dataset is composed of solute-water dimers.

 \begin{table}
  \caption{RMSE of the AVDO power-law model trained on DES15K by classes of organic compounds (kcal/mol)}
  \label{tbl:AVDO_by_group}
\centering
  \begin{tabular}{lccccc}
   & \shortstack{ Repulsive wall }
     & \shortstack{ Short-range }
    & \shortstack{ Equilibrium }
    & \shortstack{ Medium-range }
    &  MD configurations \\
  \hline
H$_2$                 & 0.9  & 0.6  & 0.2  & 0.0  & - \\
alkanes               & 1.2  & 0.8  & 0.4  & 0.1  & 0.1 \\
alkenes               & 1.9  & 1.2  & 0.4  & 0.1  & 0.1 \\
organofluorides       & 1.9  & 1.1  & 0.5  & 0.1  & 0.7 \\
organochlorides       & 2.1  & 1.0  & 0.3  & 0.1  & 0.3 \\
alkynes               & 2.1  & 1.2  & 0.3  & 0.1  & 0.1 \\
aldehydes             & 2.2  & 1.0  & 0.3  & 0.1  & 0.2 \\
ethers                & 2.7  & 1.3  & 0.5  & 0.1  & 0.3 \\
esters                & 2.7  & 1.1  & 0.2  & 0.1  & 0.2 \\
arenes                & 3.0  & 1.4  & 0.5  & 0.1  & 0.2 \\
amines                & 3.1  & 2.0  & 0.7  & 0.1  & 0.4 \\
ketones               & 3.2  & 1.3  & 0.3  & 0.1  & 0.1 \\
organobromides        & 3.2  & 1.4  & 0.1  & 0.1  & - \\
thiols                & 3.3  & 1.6  & 0.4  & 0.1  & 0.3 \\
alcohols              & 3.4  & 1.6  & 0.5  & 0.1  & 0.4 \\
thioethers            & 3.7  & 2.0  & 0.4  & 0.1  & 0.3 \\
water                 & 3.8  & 2.1  & 0.5  & 0.1  & 0.3 \\
amides                & 3.9  & 2.2  & 0.7  & 0.1  & 0.4 \\
N-heteroarenes        & 4.1  & 2.4  & 0.7  & 0.1  & 0.5 \\
carboxylic acids      & 4.3  & 2.4  & 0.8  & 0.1  & 0.3 \\
nitriles              & 5.5  & 2.8  & 0.6  & 0.0  & 0.4 \\
organophosphates      & 7.3  & 4.7  & 1.6  & 0.3  & 0.4 \\
\hline
    \end{tabular}
  \end{table}

Finally, we check the accuracy of AVDO as a function of system size by computing the exchange repulsion for acene homodimers using the experimental nearest-neighbor configuration taken from the most stable molecular crystal polymorph \cite{structures, ccdc} (the herringbone configuration, see the SI). As seen in Table \ref{tbl:acene}, the accuracy of the AVDO power-law model fit on DES15K does not degrade with system size, despite the $\pi$-$\pi$ conjugation present in these molecules, and is better than the all-electron model for larger molecules.

  \begin{table}
  \caption{Error of Exchange-Repulsion models in kcal/mol for equilibrium acene dimers (relative error is in parenthesis)}
  \label{tbl:acene}
\centering
  \begin{tabular}{lcccc}
  \hline
     &   Naphthalene 
     &  Anthracene
    &   Tetracene
    &  Pentacene \\
    \hline
All-electron   & 0.3 (5\%)  & -0.2 (2\%)  & -0.7 (6\%)   &  -0.8 (7\%)   \\
AVDO  & 0.5 (8\%)  & 0.3 (4\%)  & 0.1 (1\%)   &  0.3 (2\%)   \\
\hline
    \end{tabular}
  \end{table}
  
\section{Discussion}
The density overlap model is an empirical approach that uses the electron density as input to predict the exchange-repulsion energy. Previous models using isotropic Slater-type descriptions of the density yield density overlaps which do not correlate well enough with the exchange-repulsion energy to achieve accuracy with a few universal parameters, and so instead the density overlap needs to be weighted differently per atom-type \cite{van_vleet_2016_exch, schriber2021cliff}. While transferability to molecules in equilibrium configurations outside of the training dataset has been demonstrated for the component-based, machine-learned, intermolecular force field  (CLIFF) \cite{schriber2021cliff} and possibly for the newly developed exchange-repulsion model for the atomic multipole optimized energetics for biomolecular applications (AMOEBA) force field \cite{rackers_2019_S2R}, it is unclear to what extent these models are applicable to non-equilibrium geometries. It is also quite possible that new chemical motifs will be encountered that require fitting new atom-type parameters. Here we have shown that the overlap of anisotropic, DFT-derived valence densities of a given configuration correlates well with the exchange-repulsion energy of that configuration, such that one does not need to employ atom-type weighting. We demonstrate transferability to both new molecules and non-equilibrium conformations on a large dataset using only two universal parameters.

It is useful to compare AVDO with these transferable atom-type models. The AVDO power-law model has the same RMSE, 0.4 kcal/mol, as the AMOEBA FF for long-range dimers (0.8-0.95 $R_{eq}$) of the S101x7 dataset \cite{wang2015_penetration_s101}. For intermediate dimers of the S101x7 dataset (1.0-1.1 $R_{eq}$) AVDO has an RMSE of 1.5 kcal/mol compared to AMOEBA's RMSE of 1.0 kcal/mol. There are two caveats about this comparison. The first is that we exclude the six charged dimers in S101x7 when calculating AVDO's error. The second is that while both AVDO and AMOEBA are fit using the same dataset, S101x7, AVDO is fit and evaluated against SAPT(DFT) data, while AMOEBA is fit and evaluated against SAPT2+ data. Comparing now to the exchange-repulsion component model in CLIFF \cite{schriber2021cliff} on the S66x8 dataset \cite{s66}, the AVDO power-law model, fit on DES15K, has an RMSE of 0.6 kcal/mol, while CLIFF has an RMSE of 1.1 kcal/mol. The same caveat applies that AVDO is fit and tested on SAPT(DFT) data, while CLIFF is fit and tested on SAPT2+(3)$\delta MP2$ data. Additionally, the CLIFF exchange-repulsion component was fit as part of a global fit regularized by SAPT components, and may include error compensation with other components in CLIFF. As to the number of parameters, the AMOEBA exchange-repulsion model has 26 atom types (a total of 78 fitted parameters) and the CLIFF exchange-repulsion model has 17 atom-type parameters. Thus, our model substantially reduces the number of fitted parameters, achieves comparable accuracy, and extends to non-equilibrium configurations, albeit at a higher computational cost (discussed later).

We now comment on possible reasons for the AVDO model's transferability. As shown in eq. 3 of the SI, the SAPT zeroth-order exchange-repulsion energy depends approximately quadratically on the overlap between molecular orbitals, $S$, where $S =\sum_{ab} \int \psi_a^*(\bm{r}) \psi_b(\bm{r})d\bm{r}$, and $\psi_a$ and $\psi_b$ are molecular orbitals of molecule A and molecule B, respectively. This is the basis for the $S^2/R$ model for the exchange-repulsion energy \cite{murrell1965theory, rackers_2019_S2R}. The outer valence orbitals of many organic molecules are composed of p-orbitals or hybridized orbitals with partial p-character and contain radial nodes for 3rd row elements and beyond. The overlap of these molecular orbitals with many of the core and semi-core molecular orbitals of the other molecule, which are often s-type in nature and have fewer radial nodes, will have large cancellations/interference. However, in the density overlap model these overlaps will not cancel, as pointed out previously in other works \cite{soderhjelm_2006_review, henrichsmeyer_fink2023}. This density overestimation is element dependent. For instance, hydrogen-hydrogen contacts will have practically no molecular orbital overlap cancellation compared to hydrogen-fluorine contacts. Since heavier atoms have more orbitals and the outermost orbitals will have more nodes, the density overlap will overestimate the orbital overlap more. For this reason, we speculate that removing the overlaps of the core and semi-core orbitals lowers the density overlap in a way that systematically cancels with this overestimation.

Another possible reason is that neglecting core and semicore orbitals changes the PES of the exchange-repulsion model slightly, as shown in Fig. \ref{fig:F2_uracil_1}. The distance dependence of overlaps involving lower energy orbitals has a larger coefficient in the exponent than overlaps involving only upper valence orbitals, so it increases faster as the molecules approach each other. The distance dependence of density overlap models is slightly different than the distance dependence of the exchange-repulsion energy \cite{rackers_2019_S2R}, and this causes density overlap models to overestimate the exchange-repulsion energy in the short-range \cite{kim_1981_ovlp, misquitta2016pyridine, rackers_2019_S2R}. Removing the density overlap from semi-core molecular orbitals may reduce the distance-dependent error. Mitigating the error in the distance dependence increases the transferability of the potential because dimers will have different equilibrium distances depending on which atoms come into contact. Thus, a distance-dependent error could masquerade as poor transferability of the model.

We also note that the fitted atom-type parameters of other works approximately match the valence density definition we use in AVDO. We start by considering the atom-type parameters, $A^{exch}$, which serve as prefactors for the ISA-Slater FF exchange repulsion model \cite{van_vleet_2016_exch}. According to (eqs. 4, 14, and 25) $A^{exch} = K\sqrt{\frac{\pi}{B}}D$ where $\rho_{r\rightarrow \infty} \approx De^{-Br}$ is fit to the iterative stockholder analysis (ISA) density tail \cite{lillestolen_2009_isa} to determine the parameters $D$ and $B$, and $K$ is a coefficient to relate the density overlap to the exchange-repulsion energy. Thus $A^{exch}$ is proportional to the number of electrons that compose the tail of the density. Noting that the fitted $B$ parameters do not vary significantly and under the assumption that a universal $K$ can be found (as we have shown in this work), one sees that the fitted $A^{exch}$ atom-types approximately correspond with the number of valence electrons associated with each atom. In that study, the ratio $A_{carbon}$/$A_{hydrogen}$ for methane, ethane, and ethene was between 2.6-4 (atomic partial charges in other molecules complicate the quantitative analysis, though the trend is still apparent). The same exchange-repulsion model is used in CLIFF, which shows a similar trend that $A_{exch}$ (which they denote as $K_{exch}$) increases as one moves from carbon to fluorine. Both studies also show that $K$ remains nearly the same for sulfur, chlorine, and bromine, similar to the fact that the number of valence orbitals included in AVDO for P, S, Cl, and Br are similar to 2nd row elements. The valence density in AVDO can therefore be interpreted conservatively as an collection of implicit atom-type parameters, or alternatively, previous fitted atom-type parameters may be viewed as evidence that the valence density is an effective density for overlap models. 

The AVDO model's ability to calculate non-equilibrium configurations accurately can enable a qualitatively different application than previous models. Previous overlap models were demonstrated to be applicable to rigid-body molecular dynamics for molecules they are parameterized for \cite{van_vleet_2016_exch, van_vleet2018mastiff, janicki2023mastiff2} and can also be used to predict binding energies of new molecules in near equilibrium conformations \cite{schriber2021cliff}. Here, our model can be used to generate exchange-repulsion energies for dimers of new molecules in non-equilibrium conformations, potentially without the need of further fitting. Thus, it can be used as an essential component in new transferable force fields or intermolecular potentials that can generate synthetic data of high accuracy for training ML models.

Regarding the computational cost, in principle, the AVDO model provides the exchange-repulsion PES at the cost of a single DFT calculation of the electron density for each monomer, though here a DFT calculation is done for every configuration because we use dimer basis sets (see the SI for results of similar accuracy using smaller monomer basis sets). Recently, machine learning methods have been developed that can predict the electron density \cite{brockherde2017_ml,bogojeski2018_ml_dens, grisafi2018_transferable_ml_dens, fabrizio2019_dimer_ml_dens, bogojeski2026densnet}, with state-of-the-art models achieving an integrated absolute error in the density of 0.002 per electron on molecules from the QM9 dataset \cite{ramakrishnan2014qm9, jorgensen2022_grid_ml_dens} which were not present in the training dataset \cite{jorgensen2022_grid_ml_dens, 2022_miller_ml_dens_pnas,koker2024_ml_dens_charge3, 2024_jaakkola_neurips}. In a future publication, we will show that the AVDO model can take advantage of these recent developments so that no DFT calculation is needed. An added benefit of using a machine-learned model is that it can represent the density using a density-fitting basis \cite{brockherde2017_ml, grisafi2018_transferable_ml_dens, bogojeski2021, 2022_miller_ml_dens_pnas, rackers2023_ml_dens_water, 2024_jaakkola_neurips, bogojeski2026densnet, baerends1973DF, skylaris2000DF} and this enables the efficient calculation of the density overlap. We estimate that with a density-fit basis, the AVDO model would be on par with the cost of machine-learned potentials (about 3 orders of magnitude slower than a Born-Mayer model and models based on the overlap of Slater S-type functions \cite{van_vleet_2016_exch, schriber2021cliff}). Given this cost, we think that this model will be useful for molecular docking, geometry optimization, and the generation of datasets for machine-learning.

Finally, we consider whether AVDO can be fit on higher accuracy reference methods. We find that the parameter $K$ is relatively insensitive to changes in the density tail due to different levels of theory (see Fig. 4 in the SI). It is also known that MP2 exchange-repulsion energies are proportional to Hartree-Fock exchange-repulsion energies \cite{hodges_wheatley2000} (see also the SI) and, additionally, it has been shown that CCSD densities can be predicted by machine-learned models \cite{rackers2023_ml_dens_water}. This indicates that the model has the potential to be used/fitted on higher-accuracy ab initio methods.

\section{Conclusions}
In summary, we show that using a partial valence density instead of the all-electron density enables us to create a model for Pauli exchange-repulsion using two universal fitted parameters: the anisotropic valence density overlap (AVDO) model. The AVDO model achieves sub-kcal/mol accuracy for dimers at near equilibrium distances and is transferable across a large domain of small, neutral, closed-shell organic molecules, with accuracy that also extends to non-equilibrium configurations. Evidence suggests that the AVDO model is applicable to larger molecules as well. These advantages come at the cost of being more expensive than simpler models of exchange-repulsion. The AVDO model can be integrated in a machine-learned force field framework, where the additional cost is relatively moderate and is far outweighed by the benefit of having a model with excellent transferability. Integrating the AVDO model with an ML model for the density would make the AVDO model fully predictive. This approach has the advantage of replacing the machine-learning task of predicting dimer interaction energies with that of predicting monomer densities, a combinatorially simpler problem that substantially reduces the size of the required dataset. This work brings us one step closer to transferable, high-accuracy force fields for biological applications such as drug discovery and for generating high-accuracy, synthetic data for ML force fields.

\section*{Associated content}
\section*{Data Availability Statement}
The data underlying this study are openly available on Zenodo at https://doi.org/10.5281/zenodo.19696119.

\section*{Supporting information}
The Supporting Information is available free of charge at https://pubs.acs.org/doi/10.1021/xxxx.

The Supporting Information includes: Parameters for the AVDO model, further details for how the partial density is determined, motivating the AVDO approximation from theory, analyzing the distribution of optimal parameters by element, Boltzmann weighting, fitting AVDO to other reference methods, basis set convergence, mean signed errors for each class of molecules, a list of all molecules in the DES15K dataset by functional group, visualization of acene dimers, and a discussion about the link between the orbital overlap and density overlap models for exchange-repulsion.

\section*{Acknowledgements}
DW thanks Dr. Alston Misquitta and Prof. Reinhold Fink for helpful discussions. This research was funded by the Luxembourg National Research Fund (FNR), grant reference MBD-in-BMD C23/MS/18093472 and by the European Research Council
(ERC) Project FITMOL-101054629. The calculations presented in this paper were carried out using the HPC facilities of the University of Luxembourg \cite{UL_HPC} (see \href{http://hpc.uni.lu}{hpc.uni.lu}).

\printbibliography

\newpage

\rule{0.05in}{1.75in}%
\begin{minipage}[b][1.75in][b]{3.25in}
  \sffamily
  \frenchspacing
\centering

    \includegraphics [ width=3.25in,
     height=1.75in,
  keepaspectratio,
  trim=0 0.2in 0 0
]{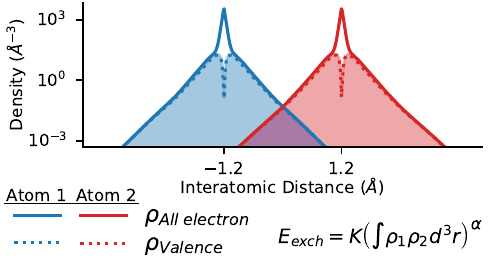}
    \label{fig:toc_figure}

\end{minipage}%
\rule{0.05in}{1.75in}

\end{document}


\maketitle
\section{Parameterization of the AVDO model}
In Table \ref{tbl:parameters} we list the parameters used for each model. All models were fit using S66 unless explicitly mentioned otherwise. Note that for linear models we can alternatively use $E = KS$, where $S= \int \rho_A \rho_B \dd[3]{r}$, as the model and this means that $K$ has the units of Bohr$^2$, a convention we often use when analyzing $K$ for linear models in this study.

  \begin{table}[htbp]
  \caption{Model Parameters. To be precise about units, let all models be of the form $E = K (S/S_0)^\alpha$ where $K$ has units of Hartree and $S_0=1$ Bohr$^{-3}$.  }
  \label{tbl:parameters}
\centering
  \begin{tabular}{lcc}
  \hline
   Model &$K$ &$\alpha$ \\ 
    \hline
    Linear Models \\
All-electron             &  32.4 & 1 \\ 
C$^2$N$^2$O$^2$          &  38.3 & 1\\
C$^2$N$^2$O$^4$          &   38.6 & 1 \\
C$^2$N$^4$O$^4$          & 41.0  & 1\\
C$^4$N$^4$O$^4$          & 58.6 & 1  \\
\hline
Power Law Models\\
All-electron             & 18.94 & 0.918  \\
All-electron - fit on DES15K & 13.83 & 0.874 \\
C$^2$N$^2$O$^2$          & 26.78 & 0.946 \\
C$^2$N$^2$O$^4$          &  28.78 & 0.955\\
C$^2$N$^4$O$^4$          &  30.36 & 0.956\\
\bf{C$^2$N$^4$O$^4$} - fit on DES15K$^*$ &  \bf{26.24} & \bf{0.937}\\
C$^4$N$^4$O$^4$          & 55.40 & 0.992 \\
\hline
    \end{tabular}
      \parbox{\linewidth}{\vspace{1ex} \centering  \small $^*$ AVDO model}
  \end{table}

Next, we show the RMSE of each model using valence densities defined by different values of $M_F$ (see eq. 2 of the main paper) for all the dimers that contain fluorine in the S101-FPSClBr dataset in Table \ref{tbl:F_compare_models}. We do the same for phosphorous in Table \ref{tbl:P_compare_models}, sulfur in Table \ref{tbl:S_compare_models}, chlorine in Table \ref{tbl:Cl_compare_models}, and bromine in Table \ref{tbl:Br_compare_models}. In bold is the model with the value of $M_X$ that yields the lowest RMSE; this determined $M_X$ in Table 2 of the main article.

  \begin{table}
  \caption{RMSE of Exchange Repulsion Models Using Different F Partial Densities (kcal/mol)}
  \label{tbl:F_compare_models}
  \centering
  \begin{tabular}{cccccc}
  \hline
    Model Parameters & $M_F$ 
    & 0.8 $R_{eq}$ &  0.9 $R_{eq}$  
    & 1.0 $R_{eq}$  & 1.1 $R_{eq}$\\
    \hline
    Linear Models \\
C$^2$N$^2$O$^2$ & F$^2$   & 5.1 & 1.2 & 0.4 & 0.1 \\
C$^2$N$^2$O$^4$ & F$^2$   & 4.7 & 0.9 & 0.2 & 0.0 \\
" " & \bf{F$^4$}  & \bf{ 1.1 } & \bf{ 0.3 } & \bf{ 0.2 } & \bf{ 0.1 } \\
C$^2$N$^4$O$^4$ & F$^2$   & 6.2 & 1.3 & 0.3 & 0.1 \\
" " & \bf{F$^4$}  & \bf{ 2.3 } & \bf{ 0.3 } & \bf{ 0.1 } & \bf{ 0.1 } \\
\hline
Power-law models \\
C$^2$N$^2$O$^2$ & F$^2$   & 5.3 & 1.8 & 0.7 & 0.3 \\
C$^2$N$^2$O$^4$ & F$^2$   & 4.8 & 1.4 & 0.5 & 0.2 \\
" " & \bf{F$^4$}  & \bf{ 1.3 } & \bf{ 0.4 } & \bf{ 0.1 } & \bf{ 0.1 } \\
C$^2$N$^4$O$^4$ & F$^2$   & 6.0 & 1.7 & 0.6 & 0.2 \\
" " & \bf{F$^4$}  & \bf{ 2.2 } & \bf{ 0.6 } & \bf{ 0.2 } & \bf{ 0.1 } \\
\hline
    \end{tabular}
  \end{table}
  

  \begin{table}
  \caption{RMSE of Exchange Repulsion Models Using Different P Partial Densities (kcal/mol)}
  \label{tbl:P_compare_models}
  \centering
  \begin{tabular}{cccccc}
  \hline
    Model Parameters & $M_P$ 
    & 0.8 $R_{eq}$ &  0.9 $R_{eq}$  
    & 1.0 $R_{eq}$  & 1.1 $R_{eq}$\\
    \hline
    Linear Models \\
C$^2$N$^2$O$^2$ & P$^{4}$   & 23.8 & 10.9 & 4.8 & 2.0 \\
" " & P$^{10}$   & 23.2 & 10.6 & 4.6 & 1.9 \\
" " & \bf{P$^{12}$}  & \bf{ 17.7 } & \bf{ 7.7 } & \bf{ 3.1 } & \bf{ 1.2 } \\
C$^2$N$^2$O$^4$ & \bf{ P$^{4}$}  & \bf{ 14.9 } & \bf{ 6.2 } & \bf{ 2.3 } & \bf{ 0.8 } \\
" " & P$^{10}$   & 4.5 & 4.4 & 3.5 & 2.6 \\
" " & P$^{12}$   & 21.3 & 14.0 & 8.9 & 5.6 \\
C$^2$N$^4$O$^4$ & P$^{4}$   & 20.7 & 9.5 & 4.1 & 1.6 \\
" " & \bf{P$^{10}$}  & \bf{ 1.6 } & \bf{ 1.6 } & \bf{ 1.8 } & \bf{ 1.6 } \\
" " & P$^{12}$   & 17.6 & 11.7 & 7.6 & 4.8 \\
\hline
Power-law models \\
C$^2$N$^2$O$^2$ & P$^{4}$   & 17.1 & 8.8 & 4.7 & 2.6 \\
" " & P$^{10}$   & 16.5 & 8.5 & 4.5 & 2.5 \\
" " & \bf{P$^{12}$}  & \bf{ 11.8 } & \bf{ 5.9 } & \bf{ 3.1 } & \bf{ 1.7 } \\
C$^2$N$^2$O$^4$ & \bf{ P$^{4}$}  & \bf{ 10.3 } & \bf{ 4.8 } & \bf{ 2.2 } & \bf{ 1.0 } \\
" " & P$^{10}$   & 7.4 & 5.0 & 3.3 & 2.0 \\
" " & P$^{12}$   & 23.0 & 14.1 & 8.5 & 5.0 \\
C$^2$N$^4$O$^4$ & P$^{4}$   & 14.5 & 7.2 & 3.6 & 1.8 \\
" " & \bf{P$^{10}$}  & \bf{ 3.7 } & \bf{ 2.9 } & \bf{ 1.9 } & \bf{ 1.2 } \\
" " & P$^{12}$   & 20.2 & 12.4 & 7.5 & 4.4 \\
\hline
    \end{tabular}
  \end{table}

  \begin{table}
  \caption{RMSE of Exchange Repulsion Models Using Different S Partial Densities (kcal/mol)}
  \label{tbl:S_compare_models}
\centering
  \begin{tabular}{cccccc}
  \hline
    Model Parameters & $M_S$ 
    & 0.8 $R_{eq}$ &  0.9 $R_{eq}$  
    & 1.0 $R_{eq}$  & 1.1 $R_{eq}$\\
    \hline
    Linear Models \\
C$^2$N$^2$O$^2$ & S$^{4}$   & 14.0 & 3.3 & 0.8 & 0.3 \\
" " & S$^{10}$   & 11.8 & 2.7 & 0.8 & 0.3 \\
" " & \bf{S$^{12}$}  & \bf{ 5.8 } & \bf{ 1.6 } & \bf{ 0.8 } & \bf{ 0.4 } \\
C$^2$N$^2$O$^4$ & \bf{ S$^{4}$}  & \bf{ 13.0 } & \bf{ 2.7 } & \bf{ 0.6 } & \bf{ 0.3 } \\
" " & S$^{10}$   & 5.2 & 1.2 & 0.8 & 0.5 \\
" " & S$^{12}$   & 4.0 & 2.3 & 1.3 & 0.7 \\
C$^2$N$^4$O$^4$ & S$^{4}$   & 17.2 & 4.1 & 0.9 & 0.2 \\
" " & \bf{S$^{10}$}  & \bf{ 8.5 } & \bf{ 1.5 } & \bf{ 0.4 } & \bf{ 0.3 } \\
" " & S$^{12}$   & 4.4 & 1.4 & 0.8 & 0.4 \\
\hline
Power-law models \\
C$^2$N$^2$O$^2$ & S$^{4}$   & 9.0 & 3.1 & 1.3 & 0.6 \\
" " & S$^{10}$   & 7.2 & 2.5 & 1.1 & 0.6 \\
" " & \bf{S$^{12}$}  & \bf{ 3.5 } & \bf{ 1.6 } & \bf{ 0.8 } & \bf{ 0.4 } \\
C$^2$N$^2$O$^4$ & S$^{4}$   & 8.7 & 2.3 & 0.6 & 0.2 \\
" " & \bf{S$^{10}$}  & \bf{ 2.2 } & \bf{ 0.7 } & \bf{ 0.4 } & \bf{ 0.2 } \\
" " & S$^{12}$   & 4.8 & 2.1 & 0.8 & 0.3 \\
C$^2$N$^4$O$^4$ & S$^{4}$   & 11.8 & 3.4 & 1.0 & 0.4 \\
" " & \bf{S$^{10}$}  & \bf{ 4.1 } & \bf{ 0.9 } & \bf{ 0.2 } & \bf{ 0.1 } \\
" " & S$^{12}$   & 2.9 & 1.2 & 0.5 & 0.2 \\
\hline
    \end{tabular}
  \end{table}

  \begin{table}
  \caption{RMSE of Exchange Repulsion Models Using Different Cl Partial Densities (kcal/mol)}
  \label{tbl:Cl_compare_models}
\centering
  \begin{tabular}{cccccc}
  \hline
    Model Parameters & $M_{Cl}$ 
    & 0.8 $R_{eq}$ &  0.9 $R_{eq}$  
    & 1.0 $R_{eq}$  & 1.1 $R_{eq}$\\
    \hline
    Linear Models \\
C$^2$N$^2$O$^2$ & \bf{ Cl$^{4}$}  & \bf{ 1.5 } & \bf{ 0.6 } & \bf{ 0.3 } & \bf{ 0.1 } \\
" " & Cl$^{10}$   & 1.6 & 0.9 & 0.4 & 0.1 \\
" " & Cl$^{12}$   & 4.2 & 1.6 & 0.6 & 0.2 \\
C$^2$N$^2$O$^4$ & \bf{ Cl$^{4}$}  & \bf{ 1.0 } & \bf{ 0.6 } & \bf{ 0.4 } & \bf{ 0.2 } \\
" " & Cl$^{10}$   & 1.7 & 1.0 & 0.5 & 0.2 \\
" " & Cl$^{12}$   & 4.5 & 1.8 & 0.7 & 0.3 \\
C$^2$N$^4$O$^4$ & \bf{ Cl$^{4}$}  & \bf{ 2.2 } & \bf{ 0.3 } & \bf{ 0.2 } & \bf{ 0.1 } \\
" " & Cl$^{10}$   & 0.6 & 0.6 & 0.3 & 0.1 \\
" " & Cl$^{12}$   & 3.2 & 1.4 & 0.5 & 0.2 \\
\hline
Power-law models \\
C$^2$N$^2$O$^2$ & Cl$^{4}$   & 1.8 & 0.7 & 0.3 & 0.1 \\
" " & \bf{Cl$^{10}$}  & \bf{ 1.6 } & \bf{ 0.6 } & \bf{ 0.2 } & \bf{ 0.1 } \\
" " & Cl$^{12}$   & 4.0 & 1.1 & 0.3 & 0.1 \\
C$^2$N$^2$O$^4$ & \bf{ Cl$^{4}$}  & \bf{ 1.1 } & \bf{ 0.4 } & \bf{ 0.1 } & \bf{ 0.1 } \\
" " & Cl$^{10}$   & 1.4 & 0.6 & 0.2 & 0.1 \\
" " & Cl$^{12}$   & 4.2 & 1.3 & 0.4 & 0.1 \\
C$^2$N$^4$O$^4$ & Cl$^{4}$   & 2.0 & 0.5 & 0.2 & 0.1 \\
" " & \bf{Cl$^{10}$}  & \bf{ 0.5 } & \bf{ 0.2 } & \bf{ 0.1 } & \bf{ 0.0 } \\
" " & Cl$^{12}$   & 3.2 & 1.0 & 0.3 & 0.1 \\
\hline
    \end{tabular}
  \end{table}

  \begin{table}
  \caption{RMSE of Exchange Repulsion Models Using Different Br Partial Densities (kcal/mol)}
  \label{tbl:Br_compare_models}
\centering
  \begin{tabular}{cccccc}
  \hline
    Model Parameters & $M_{Br}$
    & 0.8 $R_{eq}$ &  0.9 $R_{eq}$  
    & 1.0 $R_{eq}$  & 1.1 $R_{eq}$\\
    \hline
    Linear Models \\
C$^2$N$^2$O$^2$ & \bf{ Br$^{22}$}  & \bf{ 1.6 } & \bf{ 0.8 } & \bf{ 0.4 } & \bf{ 0.2 } \\
" " & Br$^{28}$   & 2.5 & 1.2 & 0.5 & 0.2 \\
" " & Br$^{30}$   & 5.9 & 2.1 & 0.7 & 0.2 \\
C$^2$N$^2$O$^4$ & \bf{ Br$^{18}$}  & \bf{ 1.7 } & \bf{ 0.7 } & \bf{ 0.4 } & \bf{ 0.2 } \\
" " & Br$^{22}$   & 1.2 & 0.9 & 0.5 & 0.2 \\
" " & Br$^{28}$   & 2.7 & 1.4 & 0.6 & 0.2 \\
" " & Br$^{30}$   & 6.1 & 2.3 & 0.8 & 0.3 \\
C$^2$N$^4$O$^4$ & Br$^{18}$   & 3.1 & 0.5 & 0.3 & 0.1 \\
" " & \bf{Br$^{22}$}  & \bf{ 1.7 } & \bf{ 0.5 } & \bf{ 0.3 } & \bf{ 0.1 } \\
" " & Br$^{28}$   & 1.1 & 0.8 & 0.4 & 0.2 \\
" " & Br$^{30}$   & 4.6 & 1.8 & 0.6 & 0.2 \\
\hline
Power-law models \\
C$^2$N$^2$O$^2$ & \bf{ Br$^{22}$}  & \bf{ 1.8 } & \bf{ 0.7 } & \bf{ 0.3 } & \bf{ 0.1 } \\
" " & Br$^{28}$   & 2.5 & 0.8 & 0.3 & 0.1 \\
" " & Br$^{30}$   & 5.7 & 1.6 & 0.4 & 0.1 \\
C$^2$N$^2$O$^4$ & Br$^{18}$   & 1.6 & 0.5 & 0.2 & 0.1 \\
" " & \bf{Br$^{22}$}  & \bf{ 1.1 } & \bf{ 0.5 } & \bf{ 0.2 } & \bf{ 0.1 } \\
" " & Br$^{28}$   & 2.5 & 0.8 & 0.3 & 0.1 \\
" " & Br$^{30}$   & 5.9 & 1.8 & 0.5 & 0.1 \\
C$^2$N$^4$O$^4$ & Br$^{18}$   & 2.6 & 0.7 & 0.2 & 0.1 \\
" " & Br$^{22}$   & 1.3 & 0.4 & 0.1 & 0.1 \\
" " & \bf{Br$^{28}$}  & \bf{ 1.2 } & \bf{ 0.4 } & \bf{ 0.1 } & \bf{ 0.0 } \\
" " & Br$^{30}$   & 4.8 & 1.4 & 0.4 & 0.1 \\
\hline
    \end{tabular}
  \end{table}

\clearpage

\section{Theory}
\begin{figure}[htbp]
\centering
\includegraphics[width=5.7cm]{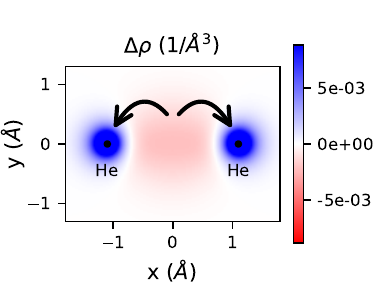}
\caption{\label{fig:density_redistribution_He2} A 2D cross-section of the redistribution of charge due to antisymmetrization for the helium homodimer. Electrons move away from the overlap region and onto each atom.}
\end{figure}

\begin{figure*}[hbtp]
    \centering
     \includegraphics[width=.95\textwidth]{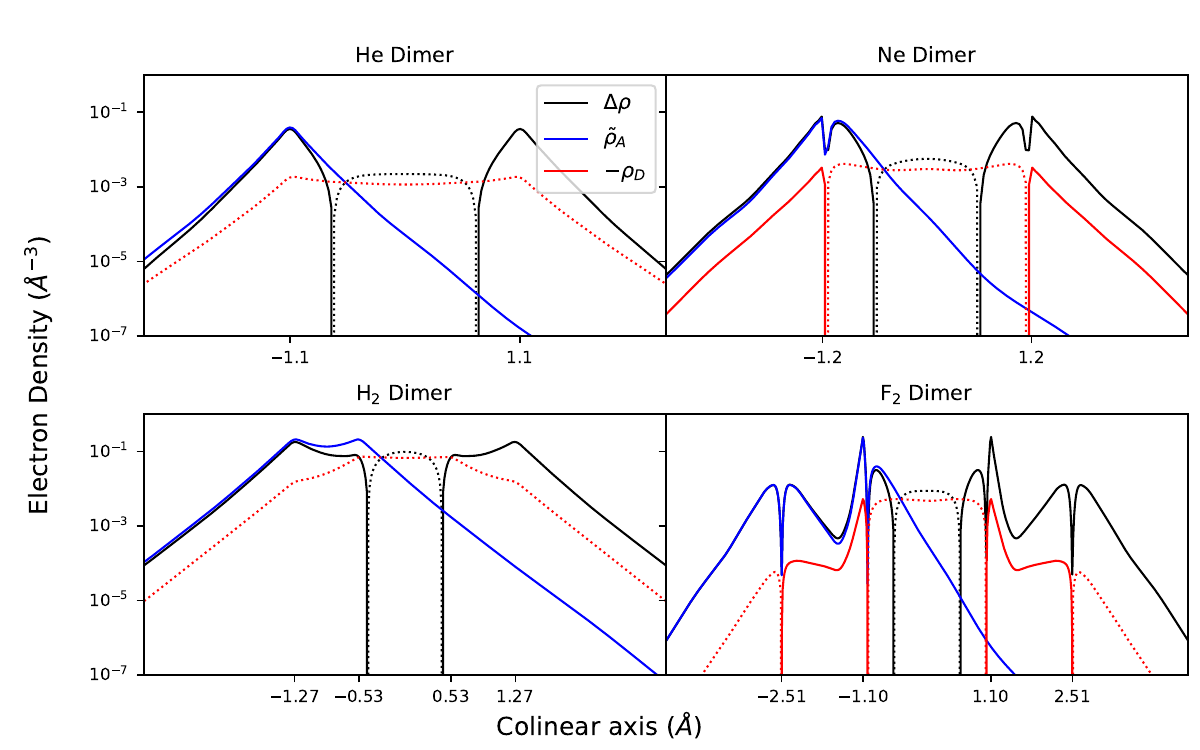}
    \hfill
    \caption{Semilog plots of the total density redistribution due to antisymmetrization $\Delta \rho$ (see Eq. 1 of the SI), the density surplus on monomer A, $\tilde{\rho_A}$, and the density deficiency $\rho_D$ (see Eq. 2 of the SI) for noble gas and diatomic dimers along the intermolecular axis. Both the H$_2$ and F$_2$ dimers are in the colinear configuration. The tick marks on the x axis indicate the positions of the atoms. Solid lines indicate positive values and dotted lines indicate negative values.}
        \label{fig:density_redistribution} 
\end{figure*}

Here we provide a brief summary of the SAPT0 exchange-repulsion energy and how it motivates the neglect of low-energy molecular orbitals. To start, we introduce the change in the electron density of two molecules in close proximity due to the antisymmetrization of their monomer wavefunctions using the $S^2$ approximation \cite{jeziorski1976first, murrell1965theory} (this is derived by taking the trace of the interaction density matrix in ref.~\cite{jeziorski1976first}):

\begin{align}
 \Delta \rho(\bm{r}) &= \bra{\Psi_A \Psi_B} \hat{\rho}(\bm{r})\mathcal{A}\ket{\Psi_A \Psi_B} - \bra{\Psi_A \Psi_B} \hat{\rho}(\bm{r})\ket{\Psi_A \Psi_B}\nonumber \\
 &\approx \tilde{\rho}_A(\bm{r})+\tilde{\rho}_B(\bm{r}) - 2\rho_D(\bm{r}),    \label{eq:delta_rho}
\end{align}
where 
\begin{align}
\tilde{\rho}_A(\bm{r}) &= 2S_{a'b}S_{ba}\psi^*_a(\bm{r}) \psi_{a'}(\bm{r})\nonumber\\
\tilde{\rho}_B(\bm{r}) &= 2S_{ab}S_{b'a}\psi_b^*(\bm{r}) \psi_{b'}(\bm{r})\nonumber\\
\rho_{D}(\bm{r}) &= 2S_{ba}\psi_a^*(\bm{r}) \psi_b(\bm{r})\nonumber\\
S_{ab} &= \bra{\psi_a}\ket{ \psi_b}.  \label{eq:rho_a_rho_d}
\end{align}
Here Einstein notation is used and $a$ and $b$ are indices for all the molecular orbitals, $\psi$, of the corresponding molecule \cite{spin_channels}. $\mathcal{A}$ is the antisymmetrizer, which is then approximated by keeping only the permutations of one electron from molecule $A$ and one electron from molecule $B$, resulting in $\Delta \rho$ containing terms only up to $S_{ab}^2$. $\rho_D$ is the depletion in the electron density in the overlap region induced by antisymmetrization. Electrons move from the depletion region onto molecule A or B and cause a surplus electron density $\tilde{\rho}_A$ and $\tilde{\rho}_B$, see Fig. \ref{fig:density_redistribution_He2}. In Fig. \ref{fig:density_redistribution} we show the redistribution of the density due to antisymmetrization for several different homodimers. $\tilde{\rho}_A$ and $\tilde{\rho}_B$ have an S-orbital character for the He and H$_2$ dimer, but have a predominantly P-orbital character for the Ne and F$_2$ dimers, as seen by the nodes in the density. This is what prompted our investigation into neglecting semicore states.

Using $\tilde{\rho}_A$, $\tilde{\rho}_B$, and $\rho_D$, we can express the first-order SAPT exchange-repulsion energy \cite{jeziorski1993sapt_expressions, moszynski1994_exch, jeziorski1976first, gordon_1996_exch, murrell1965theory} as: 

\begin{align}
\label{eq:exch_simp}
E_{exch} = & 
\int \big(\tilde{\rho}_A(\bm{r}) - \rho_D(\bm{r})\big)v_B(\bm{r})\dd[3]{\bm{r}}
+ \int \big(\tilde{\rho}_B(\bm{r}) - \rho_D^*(\bm{r})\big) v_A(\bm{r})\dd[3]{\bm{r}}\\
&-2 \sum_{ab}  \int\bigg(\psi^*_a(\bm{r})\psi_b(\bm{r})-\sum_{a'} S_{a'b}\psi_a^*(\bm{r})\psi_{a'}(\bm{r}) \bigg) \frac{1}{\abs{\bm{r}- \bm{r'}}}\bigg(\psi_b^*(\bm{r}')\psi_a(\bm{r}')- \sum_{b'}\psi_b^*(\bm{r}') \psi_{b'}(\bm{r}')S_{b'a} \bigg)\dd[3]{\bm{r}}\dd[3]{\bm{r}'}. \nonumber
\end{align}

Here $v_A$ and $v_B$ are the electrostatic potentials generated by molecules $A$ and $B$. As mentioned in the main article, DFT molecular orbitals decay as $\sim e^{-2\sqrt{-2\epsilon} r}$ \cite{misquitta2016pyridine, 2016asymptotic_dft}, where $\epsilon$ is the eigenvalue of the orbital. Thus the top-most valence orbitals of both molecules will contribute the most to $S_{ab}$ and therefore to $\tilde{\rho}_A$,  $\tilde{\rho}_B$, and $\rho_D$. Similarly, these orbitals also contribute the most to the density overlap. This also motivates the idea that low-energy molecular orbitals can be neglected in density overlap models. 

\section{Distribution of optimal $K$ parameters for each dimer}
\label{sec:parameter-distribution}
\begin{figure*}[htbp]
    \centering
     \includegraphics[width=.95\textwidth]{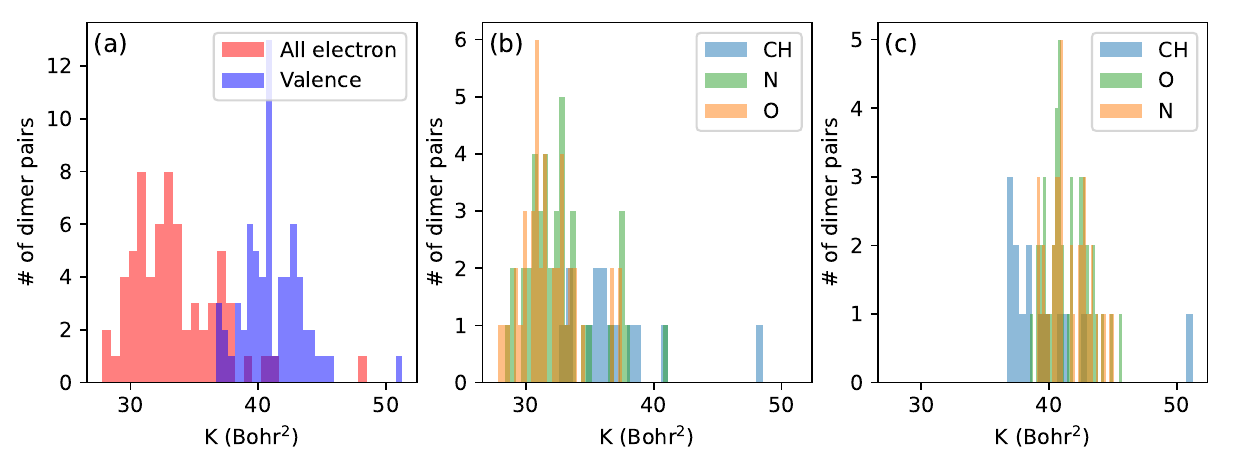}
    \hfill
    \caption{ The $K$ parameter for linear models fitted for each dimer pair separately. a) The spread of $K$ for the all-electron and valence models. The distribution of $K$ for the all-electron model (b) and the valence model (c) for hydrocarbons, molecules that contain nitrogen, and molecules that contain oxygen. The molecules which contain both nitrogen and oxygen are counted in both the oxygen and nitrogen subsets. }
    \label{fig:K_hist}
\end{figure*}

Here we investigate further how using the valence density aids transferability. There is a strong correlation between the two parameters of power-law models which complicates this analysis, so here we focus on linear models. We fit the $K$ parameter for each dimer pair of the S66 dataset individually and plot the spread of optimal $K$ values, see Fig. \ref{fig:K_hist}a. The standard deviation for $K$ for the all-electron model is 3.5 Bohr$^2$, which is 11\% of the value of $K$, while the standard deviation using the valence density defined by the AVDO model is 2.4 Bohr$^2$, which is 6\% of the value of $K$. This demonstrates how removing low-energy orbitals from the density improves the transferability of the $K$ parameter. 

In Fig. \ref{fig:K_hist}bc we investigate if molecules with different elements have different $K$ values. We show that there is a slight difference in $K$ values for molecules that contain oxygen and nitrogen for the all-electron model that is removed in the AVDO model and that hydrocarbons such as pentane, cyclopentane, etc. have $K$ values that are distinct in both groups, but closer to the other dimer $K$ values for the AVDO model.
The ethyne-ethyne dimer is the outlier with the largest $K$ value for both the all electron and AVDO model, and dimers containing ethyne have the largest $K$ values in general. Ref.~\cite{rackers_2019_S2R} removed all the dimers with ethyne when constructing the new Pauli repulsion model for AMOEBA. We keep the dimers with ethyne in the dataset because the AVDO model performs reasonably well on the other dimers with ethyne. 

\section{Boltzmann weighting}
It is expected that density overlap models will not be able to cover the entire range of distances accurately because even though the overlap density has the same exponential dependence as the exchange repulsion energy it has the wrong polynomial prefactor \cite{misquitta2016pyridine, rackers_2019_S2R}. Here, we investigate how much equilibrium configurations can be improved by using Boltzmann weighting, i.e. we minimize the sum of the absolute errors of each calculation weighted by
\[ w_i = e^{(-E_i - E_{min})/kT}\]
where $E_i$ is the CCSD(T) energy of configuration $i$, $E_{min}$ is the CCSD(T) energy of the equilibrium distance for that configuration, $k$ is the Boltzmann constant, and $T= 300K$. Note that this is not exactly Boltzmann weighting because each configuration is weighted by the minimum along it's dissociation curve as the reference, but doing this ensures that all dimer configurations are treated equally. We use the S66 dataset for the fitting procedure and linear models to facilitate the analysis. The results for these models on the S66 dataset is shown in Table \ref{tbl:Boltzmann_weighting}. Boltzmann weighting reduces the mean error for both the all-electron and AVDO models, however the error due to using a single universal $K$ keeps the RMSE at equilibrium distances from decreasing significantly. Thus, one can tradeoff between having slightly better accuracy at equilibrium at the expense of worse errors for the repulsive wall. The $K$ for the AVDO model using Boltzmann weighting is 42.7 Bohr$^2$. 

  \begin{table}[hbt]
  \caption{RMSE (Mean signed error) of Linear Exchange-Repulsion Models Using Different Weighting Schemes on S66 (kcal/mol)}
  \label{tbl:Boltzmann_weighting}
  \centering
  \begin{tabular}{lcccc}
  \hline
     &   0.8 $R_{eq}$
     & 0.9 $R_{eq}$ 
    & 1.0 $R_{eq}$
    & 1.1 $R_{eq}$
    \\
    \hline
All-electron     & 6.6 (1.7)  & 2.4 (-0.3)  & 0.9 (-0.5) & 0.5 (-0.4)  \\
All-electron (Boltzmann)  & 8.2 (3.7)  &   3.0 (0.5)   &  1.1 (-0.1) & 0.4 (-0.2) \\
AVDO & 3.6 (1.5)  &  1.1 (-0.2)  &  0.6 (-0.4) & 0.5 (-0.3) \\
AVDO (Boltzmann)& 5.3 (3.5)  & 1.6 (0.6)  &  0.5 (0.0) & 0.3 (-0.2)  \\
STD of Ref. (Mean):  &  23.9 (46.0) &   13.0 (20.6)  &  8.0 (9.5) & 4.9 (4.6) \\
\hline
    \end{tabular}
  \end{table}

\section{Fitting different reference methods}
\begin{figure}[htbp]
    \centering
     \includegraphics[width=.95\textwidth]{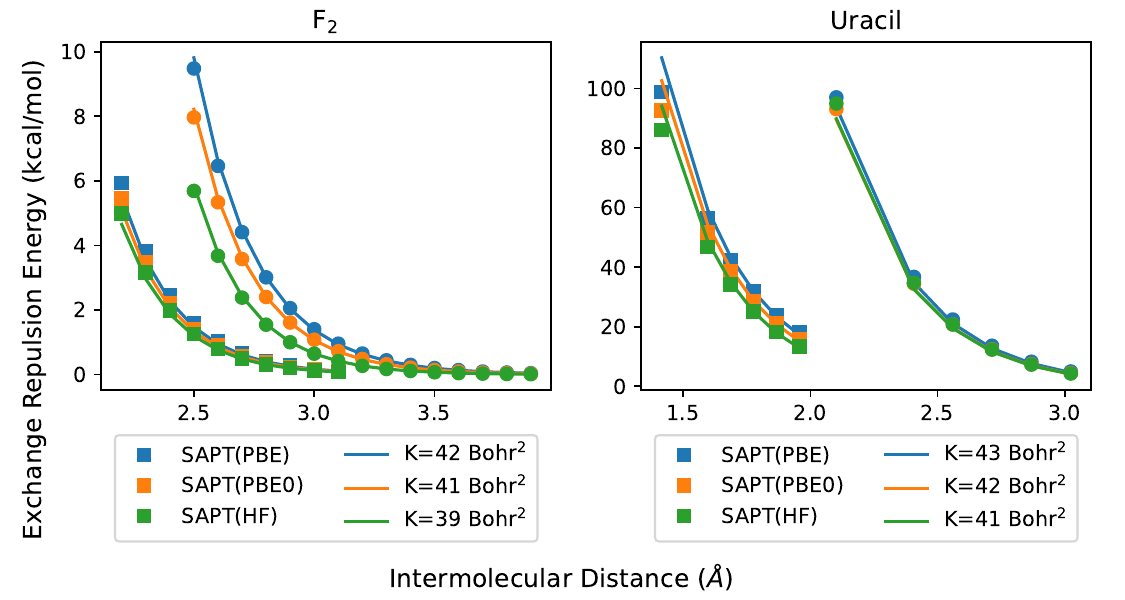}
    \hfill
    \caption{The AVDO model fitted individually for the F$_2$ dimer and the uracil dimer (lines) using several different methods to calculate the density and the SAPT exchange-repulsion energy (squares and circles, which correspond to the dimer configurations shown in Fig. 1 of the main text). The exchange-repulsion potential energy surface can vary significantly between methods, while the fitted parameter $K$ changes modestly.}
    \label{fig:F2_uracil_2}
\end{figure}

Here we investigate whether the results of the AVDO method hold for other reference methods. The density overlap strongly depends on the tails of the density. Since the density decays as $e^{-2\sqrt{2 I}}$ \cite{levy1984exact} and approximate DFT functionals give a large spread in $I$ \cite{kronik2012excitation}, the density overlap will change when using different reference methods. There are ways to correct DFT to get more accurate ionization potentials and thus have a more accurate decay of the density such as using optimally tuned range-separated hybrids \cite{kronik2012excitation} or the GRAC correction \cite{gruning2001GRAC}. However, in order to test if our results are accurate for a broad range of functionals we fit linear AVDO models using PBE0 and Hartree-Fock densities to the corresponding SAPT(DFT) and SAPT0 exchange repulsion energies. We present these results in Fig. \ref{fig:F2_uracil_2} for the F$_2$ and uracil dimers. Clearly the PES changes significantly between methods, however the density overlap also changes commensurately, such that the AVDO model remains accurate and the fitted $K$ parameter changes by a modest 5\% between methods. These results suggest that the transferability results presented for the AVDO model are applicable when using other reference methods and densities. 

\section{SAPT2 vs SAPT0 exchange repulsion}
\begin{figure}[htbp]
    \centering
     \includegraphics[width=.6\textwidth]{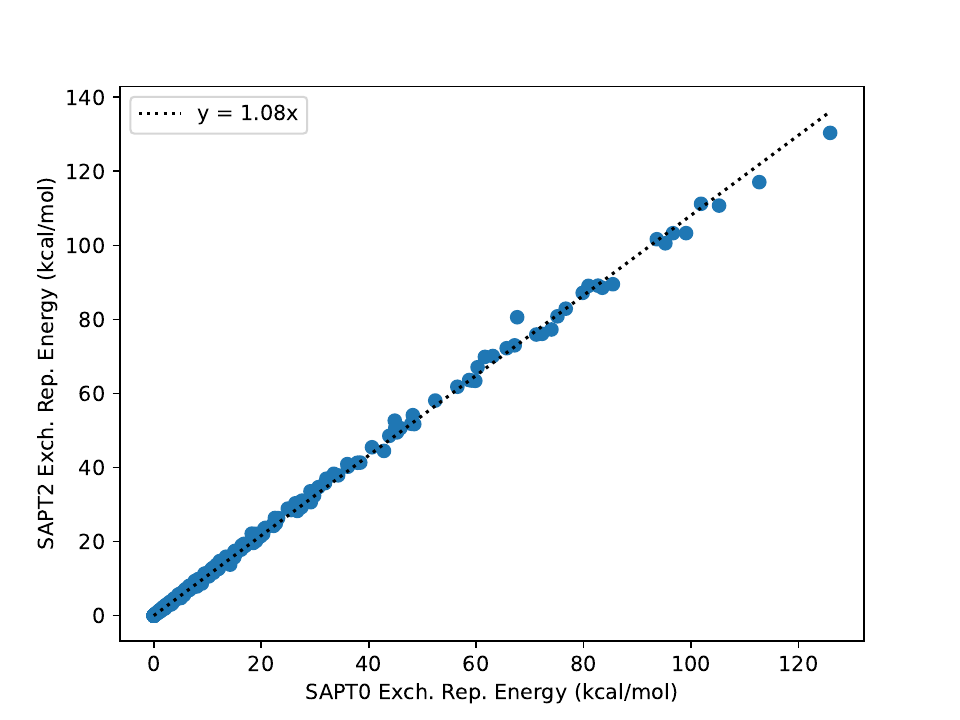}
    \hfill
    \caption{SAPT2 vs SAPT0 exchange-repulsion energies. The dotted line is a least-squares fit. }
    \label{fig:SAPT2}
\end{figure}

Here we confirm the finding of ref.~\cite{hodges_wheatley2000} that the SAPT2 exchange-repulsion energy is approximately linearly proportional to the SAPT0 exchange-repulsion energy using a larger dataset that also samples dimers at closer distances. We use data from \cite{schriber2025sapt} which consists of 4,569 calculations and covers a diverse set of systems. SAPT0 and SAPT2 exchange repulsion energies are calculated using an aug-cc-pVQZ basis. See Fig. \ref{fig:SAPT2}.

\section{Using a small monomer basis set for the density}
The density overlap for dimers farther apart than their equilibrium distances is highly dependent on the tails of the density. In Table \ref{tbl:compare_basis_sets} we quantify the effect of the basis set convergence on density overlap models by comparing linear models using the converged aug-cc-pVTZ dimer basis set to calculate the density overlap and models that use a small aug-cc-pVDZ monomer basis set. Each model is fit to reproduce the same S66 training dataset computed with SAPT(DFT) with the PBE functional on an aug-cc-pVDZ dimer basis set. We see a small, but notable, increase in error for the equilibrium, $1.0 R_{eq}$, and the medium-range, $1.1 R_{eq}$, dimers when using these under-converged densities. The $K$ parameter only slightly changes, due to needing to fit dimers at all the distances present in the dataset. This shows that the model is not very sensitive to the basis set, though performance does improve with larger basis sets.

\begin{table}
  \caption{RMSE of Linear Exchange-Repulsion Models Using Different Basis Sets on S66 (kcal/mol) }
  \label{tbl:compare_basis_sets}
  \centering
  \begin{tabular}{lccccc}
  \hline
    & 0.8 $R_{eq}$ 
     & 0.9 $R_{eq}$
    & 1.0 $R_{eq}$
    & 1.1 $R_{eq}$
    & $K$ (Bohr$^{2}$)\\
    \hline
All-electron dimer basis set     & 6.6  & 2.4  & 0.9   &  0.5  & 32.4  \\     
All-electron monomer basis set   & 6.3  & 2.4  & 1.1  &  0.7  & 32.2  \\    
AVDO dimer basis set             & 3.6  & 1.1  & 0.6   & 0.5 & 41.0 \\
AVDO monomer basis set           & 3.7  & 1.2  & 0.8  & 0.6 & 40.5 \\
\hline
    \end{tabular}
  \end{table}

\section{Mean signed errors of the AVDO model by class of organic compounds}
In Table \ref{tbl:AVDO_signed_err_by_group} we show the mean signed error of the AVDO model by class of organic molecules. This shows that there are large systematic errors for organophosphates, nitriles, and organobromides.

\begin{table}
  \caption{Mean signed error of the AVDO power-law model trained on DES15K by classes of organic compounds (kcal/mol)}
  \label{tbl:AVDO_signed_err_by_group}
\centering
  \begin{tabular}{lccccc}
   & \shortstack{ Repulsive wall }
     & \shortstack{ Short-range }
    & \shortstack{ Equilibrium }
    & \shortstack{ Medium-range }
    &  MD configurations \\
  \hline
H$_2$                &   -0.1 &   -0.1 &    0.0 &    0.0 &     - \\
alkanes              &    0.3 &    0.4 &    0.3 &    0.1 &     0.1 \\
alkenes              &   -0.7 &   -0.3 &    0.1 &    0.1 &     0.0 \\
organofluorides      &    1.1 &    0.8 &    0.4 &    0.1 &     0.4 \\
organochlorides      &   -0.8 &   -0.2 &    0.2 &    0.1 &    -0.1 \\
alkynes              &   -0.9 &   -0.5 &   -0.0 &    0.0 &    -0.0 \\
aldehydes            &   -0.8 &   -0.5 &    0.0 &    0.0 &     0.1 \\
ethers               &    1.6 &    0.8 &    0.4 &    0.1 &     0.2 \\
esters               &   -0.8 &   -0.6 &    0.0 &    0.1 &     0.1 \\
arenes               &    0.0 &   -0.0 &    0.2 &    0.1 &     0.0 \\
amines               &   -1.2 &   -1.1 &   -0.2 &    0.0 &    -0.1 \\
ketones              &   -1.5 &   -0.7 &    0.1 &    0.1 &     0.0 \\
organobromides       &   -2.8 &   -1.2 &    0.0 &    0.1 &     - \\
thiols               &   -1.1 &   -0.5 &    0.1 &    0.1 &    -0.1 \\
alcohols             &    0.1 &   -0.4 &    0.1 &    0.1 &     0.3 \\
thioethers           &   -2.5 &   -1.1 &    0.0 &    0.1 &    -0.0 \\
water                &   -2.0 &   -1.3 &   -0.2 &    0.0 &    -0.0 \\
amides               &   -1.0 &   -1.2 &   -0.2 &    0.0 &     0.0 \\
N-heteroarenes       &   -1.9 &   -1.6 &   -0.3 &    0.0 &    -0.3 \\
carboxylic acids     &   -0.3 &   -1.2 &   -0.4 &   -0.0 &     0.0 \\
nitriles             &   -2.9 &   -2.0 &   -0.4 &    0.0 &    -0.3 \\
organophosphates     &   -5.9 &   -4.0 &   -1.3 &   -0.1 &    -0.2 \\
\hline
    \end{tabular}
  \end{table}

For convenience, we also list the smiles strings by class of all molecules that appear in the subset of DES15K used in this study. The smiles strings that are in italics indicate molecules that are only in the optimized geometry dataset. All other strings are in both the optimized geometry dataset and the MD dataset.

\underline{alcohols}

CCCO, CCO, CC(O)C, CO, OC1CCCC1, OC1CCCCC1, OCCCCO, OCCCO, OCCO

\underline{aldehydes}

CC=O, C=O

\underline{alkanes}

C1CCCC1, C1CCCCC1, C, CC, CCC, CC(C)C, CCCC, CC(C)(C)C, CCCCC, CCCCCC

\underline{alkenes}

C=C, CC=C, CC=CC, CC(=C)C, CC=C(C)C, CC(=C(C)C)C

\underline{alkynes}

CC\#CC, CC\#C, C\#C, \textit{CCC\#C}

\underline{amides}

CC(=O)N, CC(=O)N(C)C, CNC=O, CNC(=O)C, NC=O, O=CN(C)C

\underline{amines}

C1CCCN1, C1CCCNC1, CCN, CCN(C)C, CN, CNC, CN(C)C, CNCC, N

\underline{arenes}

c1ccccc1, Cc1ccccc1, Oc1ccccc1

\underline{carboxylic acids}

CC(=O)O, OC=O

\underline{esters}

COC(=O)C, COC=O

\underline{ethers}

C1CCCO1, C1CCCOC1, C1CCOCO1, C1OCCO1, CCCOC, CCOCC, COCC, COC, COCOC, O1CCOCC1, O1COCOC1

\underline{H$_2$}

\textit{[H][H]}

\underline{ketones}

CC(=O)C

\underline{N-heteroarenes}

c1cccnc1, c1ccncn1, n1ccncc1, c1ccc2c(c1)[nH]cc2, c1ccc[nH]1, c1ncc[nH]1, Cc1cnc[nH]1, Cc1c[nH]cn1

\underline{nitriles}

CC\#N, C\#N, \textit{CCC\#N}

\underline{organobromides}

\textit{BrCBr}, \textit{BrCCBr}, \textit{CBr}, \textit{CC(Br)Br}

\underline{organochlorides}

CC(Cl)Cl, CCCl, CCl, Clc1ccc(cc1)Cl, Clc1ccccc1, ClCCCl, ClCCl, \textit{CCCCl}, \textit{CC(Cl)(Cl)Cl}, \textit{ClC(Cl)Cl}

\underline{organofluorides}

CC(F)F, CCF, CF, Fc1ccc(cc1)F, Fc1ccccc1, FCCF, FCF, \textit{CCC(F)(F)F}, \textit{CCC(F)F}, \textit{CCCF}, \textit{CC(F)(F)F}, \textit{Fc1cc(F)cc(c1)F}, \textit{FC(C(F)(F)F)(F)F}, \textit{FC(F)F}

\underline{organophosphates}

COP(=O)(OC)OC, COP(=O)(OC)O, \textit{COP(=O)(O)O}, \textit{OP(=O)(O)O}

\underline{thioethers}

C1CCCS1, C1CCCSC1, C1CCSCS1, C1CCSSC1, C1CSSC1, C1SCCS1, CCSCC, CCSSCC, CCSSC, CSCC, CSCSC, CSC, CSSC, S1CCSCC1, S1CSCSC1

\underline{thiols}

CCSS, CCS, CSS, CS, SS, S

\underline{water}

O
  
\section{Acene Dimers}
\begin{figure}[htbp]
    \centering
     \includegraphics[width=.95\textwidth]{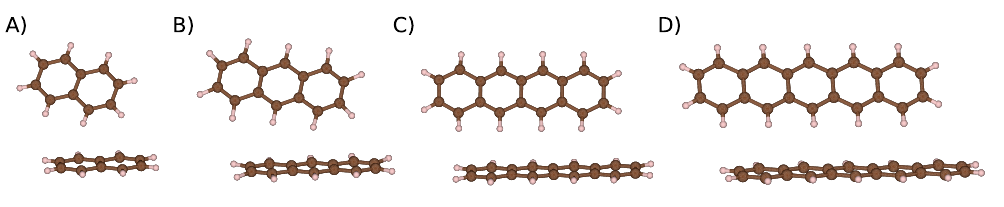}
    \hfill
    \caption{A) Naphthalene dimer, B) Anthracene dimer, C) Tetracene dimer, D) Pentacene dimer used in Table 6 of the main text.}
    \label{fig:acene}
\end{figure}

In Fig. \ref{fig:acene} we show the configuration of the acene dimers used in Table III of the main text.

\section{Connection between Orbital Overlap and Density Overlap models and explanation of the magnitude of the $K$ parameter}
We will focus on linear models for this discussion to facilitate the analysis. We first consider noble gas homodimers starting from the orbital overlap model \cite{murrell1965theory,rackers_2019_S2R}, 
\[E_{\textrm{exch}}= \frac{K_2}{r_{AB}} \sum_{ab} \big| \langle \psi_a \mid \psi_b \rangle \big|^2,\]
where $r_{AB}$ is the interatomic distance and $K_2$ is a unitless parameter that is nearly equal to one \cite{murrell1965theory}. This approximation is motivated by the form of the SAPT exchange-repulsion energy (see eq. \ref{eq:exch_simp}). Using the arithmetic mean - geometric mean inequality (assuming that the wavefunctions are real),
\begin{equation}
\label{eq:inequality}
 \bigg(\int \psi_a \psi_b \textrm{d}^3r\bigg)^2 \le V\int
\psi_a^2 \psi_b^2\textrm{d}^3r ,
\end{equation}
where $V$ is the volume of the integral, we find that
\begin{equation}
\label{eq:S_vs_S_rho}
\frac{K_2}{r_{AB}}\sum_{ab}\big| \langle \psi_a \mid \psi_b\rangle\big|^2 \le \frac{K_2V}{r_{AB}}\int \rho_A \rho_B\textrm{d}^3r. 
\end{equation}
Thus the arithmetic mean-geometric mean inequality provides a way to compare the two models. Eq. \ref{eq:S_vs_S_rho} also provides an explanation for the magnitude of the value of the $K$ parameter, which is 32.4 Bohr$^2$ for the linear all-electron density overlap model. We approximate the volume $V$ as a cylinder between two atoms with a radius defined by the atom's van der Waals radius, $V = \pi r_{vdW}^2 r_{AB}$. Since $K \approx K_2 V/r_{AB}$ and $K_2 \approx 1$, $K$ is approximately the cross-sectional area of an atom. Using a van der Waals radius of 2.8 Bohr yields $K\approx 25$ Bohr$^2$. This explanation can be extended to molecules by starting with localized Boys orbitals \cite{boys1960localized} rather than molecular orbitals in Eq. \ref{eq:inequality}. The resulting inequality is the same as Eq. \ref{eq:S_vs_S_rho}, where $V$ can still be approximated as a cylinder between two atoms, rather than the volume of the overlap region between the two molecules.

\printbibliography